# LLM Signals in Funded Grants: Evidence from International Funding Agencies

M.Z. Naser, PhD, PE
Artificial Intelligence Research Institute for Science and Engineering, Clemson University, USA
E-mail: mznaser@clemson.edu, Website: www.mznaser.com

**Absract**

The release of ChatGPT in November 2022 introduced a writing tool of unprecedented fluency into the daily routines of researchers across the sciences. Prior work has measured what follows in journal abstracts and in peer reviews by documenting an upward shift in the frequency of words and short phrases that large language models (LLMs) tend to overproduce. However, a much less-examined question concerns whether the same shift extends to grant proposals, the documents through which researchers compete for public funding, and, in particular, to the subset of proposals that succeed. Such a distinction is important to note because funded grants represent decisions to allocate public research resources, and because the proposing population spans every scientific discipline in which a country supports research. This **brief report** draws on an analysis of 221,425 funded grant abstracts from three major science-funding agencies in two countries: the U.S. National Science Foundation (NSF), the U.S. National Institutes of Health (NIH), and UK Research and Innovation (UKRI). The window covers late 2017 through mid-2026, providing roughly 5 years of pre-ChatGPT baseline and 3.5 years of post-release observation in each corpus.

*Keywords*: Large language models, Research funding, Grant proposals, Scientific communication.

## A measurable rise across three funders

The framework applied here counts the frequency of a 68-element lexical marker set established in prior post-ChatGPT work [1–4] and converts the result to a rate per 100,000 word tokens. Across the three funding agencies, the direction is the same, and the magnitudes are comparable. At NSF (n = 96,020), the mean marker rate rose from 274.7 per 100,000 tokens in the pre-period ChatGPT to 417.1 in the post-period, an increase of 51.9% with Cohen's $d$ of 0.348. NIH (n = 80,496) rose from 266.6 to 339.3, an increase of 27.3% with $d$ of 0.210. UKRI (n = 44,909) rose from 224.1 to 344.3, an increase of 53.6% with $d$ of 0.280. All three distributional shifts are highly significant under Mann-Whitney U testing against the pre-period distribution.

Figure 1 displays the three series jointly. The NSF and UKRI marker rates in the most recent quarters of 2026 are well below their 2024–2025 peaks. The NIH marker rate has continued to climb in the same window and now sits at or above the levels the NSF and UKRI series reached at their respective peaks. The three agencies, which moved roughly in parallel through the pre-ChatGPT window and through the initial post-release rise, are no longer moving in parallel.

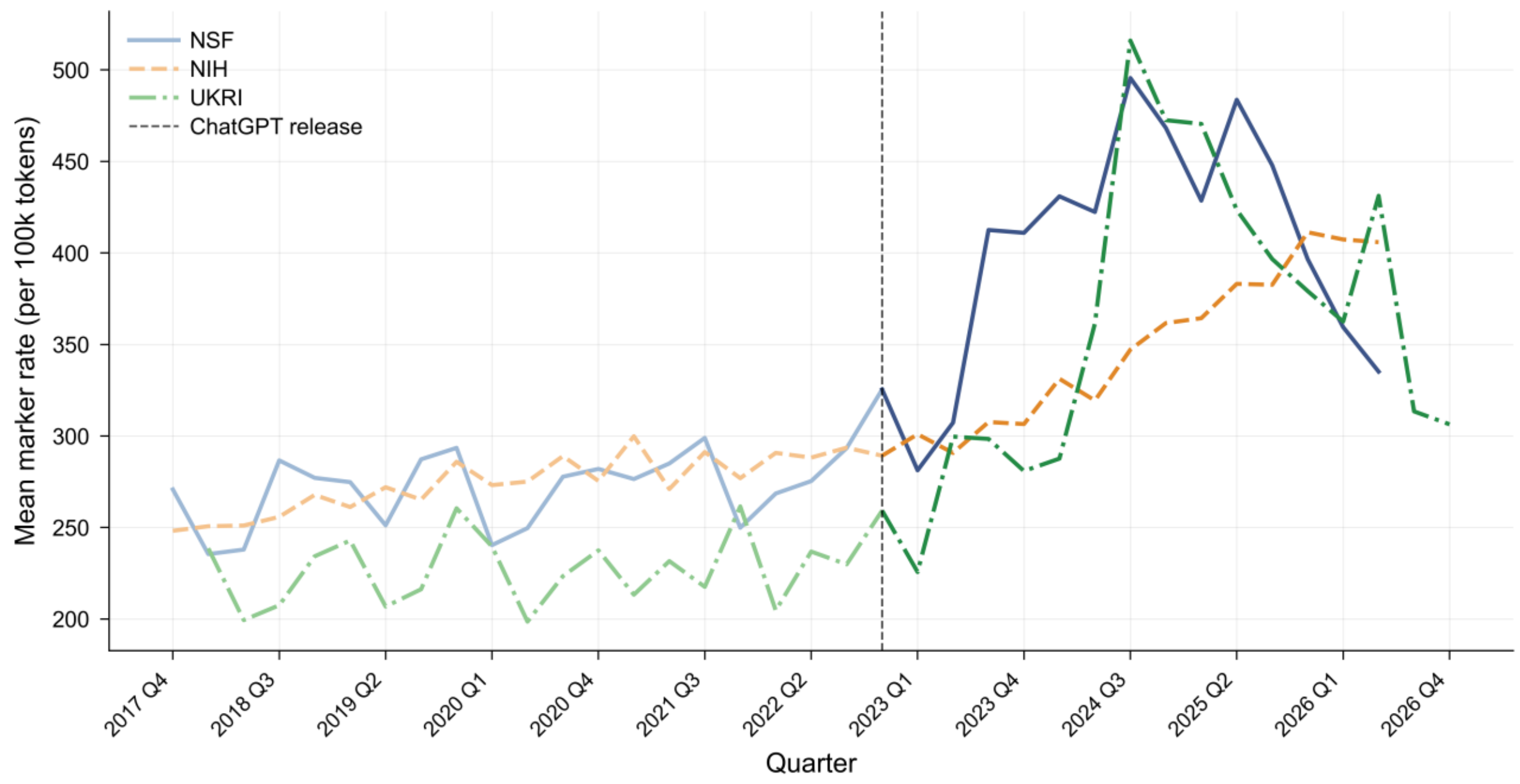

Fig. 1 Mean marker rate across NSF, NIH, and UKRI pre- and post-release of ChatGPT

The key percentages translate further into an estimate of the fraction of individual abstracts whose marker density is most consistent with substantial LLM assistance. Taking the pre-ChatGPT 95th and 99th percentile marker rates within each agency as the baseline for naturally high marker density yields an excess in the post-period ranging from 4.0% to 8.4% at NSF, 1.4% to 3.7% at NIH, and 2.8% to 6.8% at UKRI. The pre-ChatGPT distribution functions here as a model of how marker-rich abstracts could occur under writing-style variation alone [5]. The post-period excess above each upper-tail percentile is treated as a lower bound on substantial-LLM-assistance prevalence under the assumption that the underlying writing-style distribution would otherwise have remained stable. We report that roughly one in every twelve to twenty-five funded NSF proposals in the post-ChatGPT period sits in the marker-density region that was rare under pre-period writing alone, and the corresponding band at NIH is one in twenty-seven to seventy-one. The UKRI numbers reflect roughly one in every fifteen to thirty-six funded proposals post-ChatGPT.

**Where the rise concentrates**

The aggregate shift documented above masks substantial internal variation, with the rise unevenly distributed across each agency. For example, at NSF, the largest increases come from the directorates spanning technology translation, computer and information science, and the mathematical and physical sciences, with engineering and education following close behind. The smallest come from the geosciences and the social, behavioral, and economic sciences. At UKRI, a similar ordering appears in a different vocabulary. Innovate UK and the Biotechnology and Biological Sciences Research Council show the largest increases. The Arts and Humanities Research Council and the Natural Environment Research Council show the smallest. At NIH, the variation across institutes and centers is narrower, with the National Institute of General Medical Sciences and the National Cancer Institute leading and the National Institute on Aging trailing.

The cross-agency consistency of this ordering warrants closer examination (see Fig. 2). In NSF and UKRI, the units closest to applied technology and translational research exhibit the largest post-ChatGPT lexical shifts. The units closest to environmental sciences and the humanities show the smallest. NIH, which funds biomedicine almost exclusively and lacks a humanities-or-geoscience analog, falls in the middle range as an agency and shows narrower internal variation than the other two.

**What changed and what did not**

The marker-rate increases do not appear to reflect a more general shift in how grant abstracts are structured. Three syntactic metrics computed in parallel, namely average sentence length, total abstract length, and sentence count per abstract, remain essentially flat across all three agencies between the pre- and post-ChatGPT periods. The pre-to-post differences in these metrics are small in absolute terms and do not show a consistent direction across the three datasets.

This implies that if post-ChatGPT abstracts were systematically longer or more complex in sentence structure, the marker-rate increase could, in principle, be an artifact of having more text in which markers can appear. However, the flatness of the syntactic metrics rules out that confound. The observed shift is specifically lexical, with the same number of sentences, roughly the same length, but with a measurably higher frequency of words and short phrases that LLMs tend to overproduce. Whatever process is producing the shift, it operates at the level of word selection rather than at the level of structural composition.

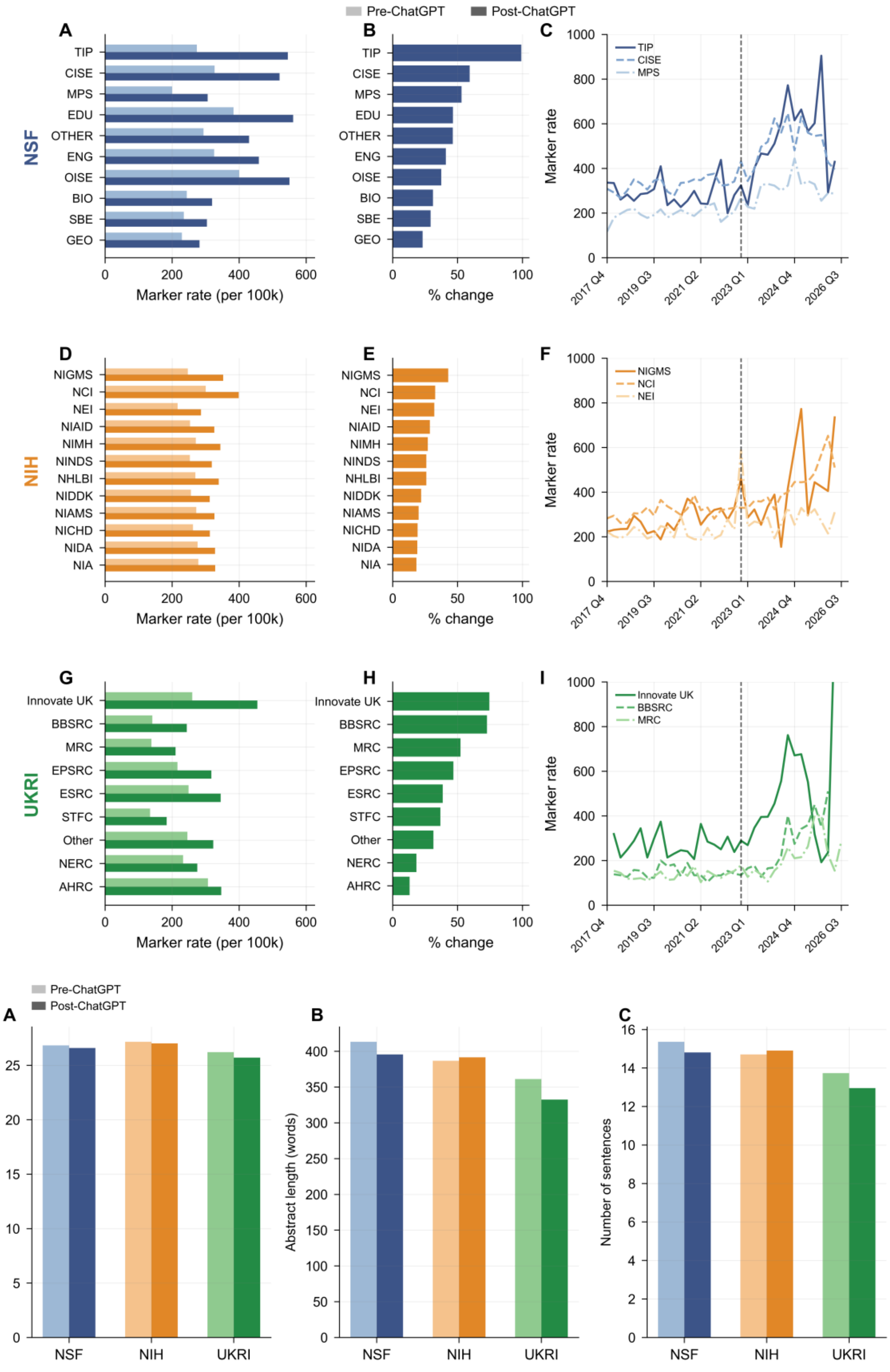

Fig. 2 Close examination of units within each funding agency

**A recent inflection that warrants continued monitoring**

The quarterly trajectory of the marker rate across agencies reveals a pattern not evident in the pooled pre-versus-post comparison (see Fig. 1). From the second quarter of 2023 onward, the marker rate across all three agencies rose steeply, reaching the peaks that drive the headline numbers reported in the previous section. The trajectories then diverge.

The interpretation of this divergence requires caution. Six to eight quarters of post-peak data in two of three agencies is insufficient to determine whether the recent decline at NSF and UKRI reflects a stable inflection, a temporary fluctuation, or an artifact of the most recent quarters being incompletely represented in the harvest at the time of writing. Several mechanisms could produce a population-level lexical decline without a corresponding change in underlying LLM use, including improvements in newer-generation language models that produce text with fewer of the markers established by earlier studies, shifts in the disciplinary mix of recent awards, changes in agency editorial practice, and genuine shifts in how researchers compose proposals when they are aware that linguistic markers are being studied. The data presented here do not separate these mechanisms. It may take a few more years of post-peak observation to determine which combination of factors best aligns with the trajectory.

**What can and cannot be concluded**

Across three major funding agencies in two countries, marker rates rose by 27% to 54% after the public release of ChatGPT, with effect sizes in the small-to-medium range and with sample sizes in the tens of thousands per agency. The shift is consistent across agencies and consistent in its disciplinary signature within NSF and UKRI, and does not seem to be explained by changes in abstract length or sentence structure. What cannot yet be concluded from the available data is whether the lexical signal continues to track underlying LLM use as the technology and its surrounding practice evolve. The recent quarterly inflection at NSF and UKRI is real in the sense that it appears in the data. What it means is unclear. The instrument that detects LLM-influenced writing depends on the joint state of three things over which researchers have no control, namely, which models are in use at any given time, which words and phrases those models tend to overproduce, and what writers know about the markers being studied. A change in any of these three would change the signal without necessarily changing the behavior the signal was originally designed to detect [6]. The implication for any policy that draws on linguistic markers as evidence of LLM use is direct. The marker rate is a contemporary measurement of a moving phenomenon, not a stable instrument [7]. Its meaning depends on a context that the measurement alone cannot reconstruct from cross-sectional data.

**What the next few years will “likely” determine**

If marker rates in NSF and UKRI stabilize or resume rising over the next few years, the dominant explanation will be that the post-peak quarters in the present analysis reflect transient noise or sampling effects. If marker rates continue to fall, the question of mechanism becomes pressing, with model evolution, writer adaptation, and policy environment as candidate accounts that the data presented here cannot separate. If NIH eventually follows the NSF and UKRI trajectories with the same approximate lag, that pattern itself becomes evidence of how awareness of detection methodology propagates across different scientific communities. that said, it is unlikely that these futures can be selected from the present data. All of them are observable in the data that will accumulate.

## Data availability statement
Data is available on request from the author.

## Conflict of interest
The author declares no conflict of interest.

## Funding declaration
None.

## Authors' contributions
There is one author in this paper.

## Acknowledgements
None.

## References
[1] W. Liang, Y. Zhang, Z. Wu, H. Lepp, W. Ji, X. Zhao, H. Cao, S. Liu, S. He, Z. Huang, D. Yang, C. Potts, C.D. Manning, J.Y. Zou, Mapping the Increasing Use of LLMs in Scientific Papers, (2024). https://arxiv.org/pdf/2404.01268 (accessed January 18, 2026).

[2] D. Kobak, R. González-Márquez, E.Á. Horvát, J. Lause, Delving into LLM-assisted writing in biomedical publications through excess vocabulary, Sci. Adv. . (2025). https://doi.org/10.1126/sciadv.adt3813.

[3] M. Geng, R. Trotta, Is ChatGPT Transforming Academics' Writing Style?, (2024). https://arxiv.org/pdf/2404.08627 (accessed January 17, 2026).

[4] Z. McCreery, M.Z. Naser, LLM-Assisted writing in engineering is associated with higher citations as evidenced from 1.17 million papers, J. Informetr. 20 (2026) 101842. https://doi.org/10.1016/J.JOI.2026.101842.

[5] J.E. Casal, M. Kessler, Can linguists distinguish between ChatGPT/AI and human writing? A study of research ethics and academic publishing, Res. Methods Appl. Linguist. (2023). https://doi.org/https://doi.org/10.1016/j.rmal.2023.100068.

[6] M. Lin, D. Liu, Intersubjectivity as the distinguishing feature or common ground: A contrastive study between human-written abstracts and LLM-generated abstracts, Res. Methods Appl. Linguist. (2026). https://doi.org/https://doi.org/10.1016/j.rmal.2026.100297.

[7] V.S. Sadasivan, A. Kumar, S. Balasubramanian, W. Wang, S. Feizi, Can AI-Generated Text be Reliably Detected? Stress Testing AI Text Detectors Under Various Attacks, Trans. Mach. Learn. Res. (2025).